\documentclass[aps, prst, reprint]{revtex4-1}

\usepackage{graphicx, epsfig, array, subfigure, hyperref, epstopdf}

\begin{document}

\title{Ultra-high quality factors in superconducting niobium cavities in ambient magnetic fields up to 190~mG}
\thanks{This work was supported by the US Department of Energy, Office of Nuclear Physics.}%

\author{A. Romanenko} 
\email{aroman@fnal.gov}
\author{A. Grassellino}
\author{A. C. Crawford}
\author{D. A. Sergatskov}
\author{O. Melnychuk}
\affiliation{Fermi National Accelerator Laboratory, Batavia, IL 60510, USA }

\date{\today}

\begin{abstract}
Ambient magnetic field, if trapped in the penetration depth, leads to the residual resistance and therefore sets the limit for the achievable quality factors in superconducting niobium resonators for particle accelerators. Here we show that a complete expulsion of the magnetic flux can be performed and leads to: 1) record quality factors $Q > 2\times10^{11}$ up to accelerating gradient of 22~MV/m; 2) $Q\sim3\times10^{10}$ at 2~K and 16~MV/m in up to 190~mG magnetic fields. This is achieved by large thermal gradients at the normal/superconducting phase front during the cooldown. Our findings open up a way to ultra-high quality factors at low temperatures and show an alternative to the sophisticated magnetic shielding implemented in modern superconducting accelerators. 
\end{abstract}


\maketitle

\begin{figure*}[htb]
 \includegraphics[width=\textwidth]{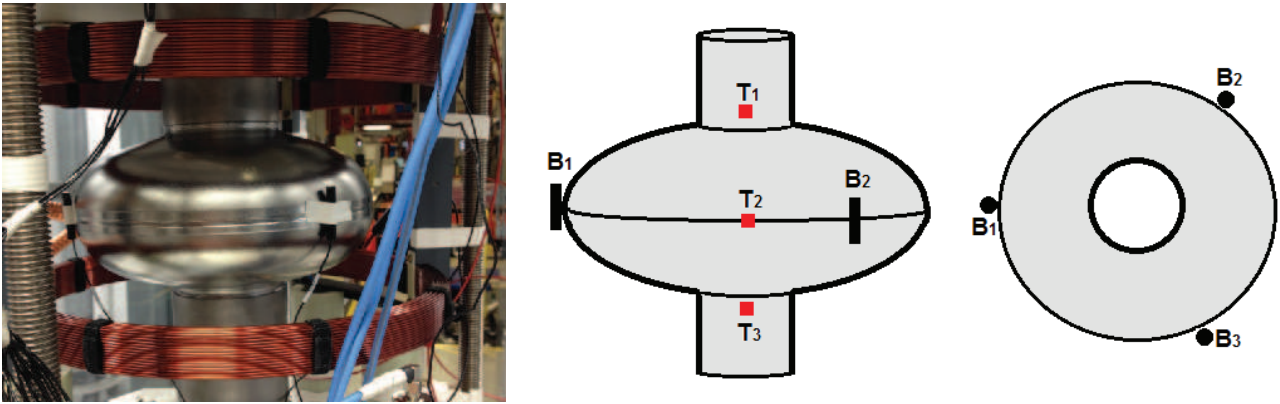}
\caption{\label{fig:Schematic}Picture of the setup and schematic of the magnetic and temperature probes mounting used for the measurements.}
\end{figure*}

Trapped magnetic flux increases the microwave surface resistance $R_\mathrm{s}$ of superconducting radio frequency (SRF) cavities by contributing to the residual (temperature independent) component $R_\mathrm{res}$ of $R_\mathrm{s}$. Its contribution is thought to come from the normal conducting cores of the trapped fluxoids~\cite{HasanBook}.  As the strongly temperature dependent $R_\mathrm{BCS}(T)$ is exponentially vanishing at lower temperatures, the minimum achievable value of $R_\mathrm{s}$ remains limited by $R_\mathrm{res}$, thereby setting the limit on the achievable quality factor $Q \propto 1/R_\mathrm{s}$. For fine grain ($\sim$50~$\mu$m) size niobium used to manufacture the majority of SRF cavities the previous understanding was that close to 100\% of the ambient magnetic field gets trapped during the transition to superconducting state. Magnetic shielding to lower the magnetic field amplitude at cavity walls was considered the only option to avoid the increased $R_\mathrm{res}$ and increased wall dissipation, and has therefore been implemented as the essential part of all SRF accelerators.

It was discovered at Helmholtz Zentrum Berlin (HZB) that the residual resistance of niobium cavities can be affected by the cooling dynamics around niobium $T_\mathrm{c}=9.25$~K~\cite{Kugeler_SRF_2009, Vogt_PRST_2013}. The cavity used for HZB studies was dressed - meaning that it had the titanium helium vessel welded on it - and had no temperature or magnetic field measurement probes on the cavity walls. Based on the readings of the temperature sensors on the beam tubes outside of helium vessel the effect was attributed to the additional magnetic field generated by thermal currents flowing through the thermocouple loop created by the cavity and titanium vessel, which gets trapped during the cooldown through $T_\mathrm{c}$. Theoretical analysis showed~\cite{Crawford_arXiv_2014} that broken current flow symmetry is required in order for this contribution to be non-negligible.

A different physical mechanism was subsequently discovered at Fermilab~\cite{Romanenko_JAP_2014} by mounting the fluxgate magnetometers and temperature sensors directly on the walls of both bare and dressed cavities. The residual resistance was demonstrated to be tracking the changes in the trapped fraction of the ambient magnetic field, and the better expulsion/lower $R_\mathrm{res}$ to correspond to the larger temperature gradients at the normal/superconducting transition front during the cooling through $T_\mathrm{c}$.  This new effect suggested that much higher fields than previously thought could in principle be expelled using high enough thermal gradients at $T_\mathrm{c}$. 

In this paper we report the discoveries of: 1) full flux expulsion leading to record $Q$ values of $>2\times10^{11}$ up to 21~MV/m; 2) almost complete flux expulsion leading to $Q\sim3\times10^{10}$ at 2~K and $E_\mathrm{acc}=16$~MV/m even in high magnetic fields of $B\leq190$~mG attainable with little or no magnetic shielding. Detailed temperature and magnetic field measurements show that the determining parameter is the temperature gradient $dT/dx$ at the normal/superconducting phase front and reveal its threshold values required for efficient flux expulsion of about 0.1-0.2~K/cm.

We used a 1.3~GHz single cell TESLA shaped cavity for our studies, which was prepared by nitrogen doping~\cite{Grassellino_SUST_2013}. Among the cavity preparation procedures, nitrogen doped cavities possess highest quality factors and are the most sensitive to the trapped flux, thus making them the ideal tool to study the flux expulsion. All the measurements were performed at Fermilab vertical testing facility, which has magnetic shielding with the ambient fields typically reduced down to $<$5~mG values. In order to control the magnetic field, Helmholtz coils were assembled around the cavity. Three single-axis Bartington Mag-01H cryogenic fluxgate magnetometers were mounted around the equator with $\sim$120$^\circ$ spacing to measure the magnetic field along the cavity axis (vertical) direction. Three Cernox temperature sensors were mounted as follows: one at the top iris, one at the equator, and one at the bottom iris. The picture of the setup and the schematic of the probe mounting is shown in Fig.~\ref{fig:Schematic}. We define $B_\mathrm{avg}=(B_1+B_2+B_3)/3$.

The cooldowns were performed under different ambient magnetic fields ranging between 2~mG and 190~mG and from different starting temperatures ranging from 300~K to 12~K. The $Q(E)$ measurements were performed at 2~K and the lowest achievable temperature of 1.5~K for most of the measurements. 

\begin{figure}[htb]
 \includegraphics[width=\linewidth]{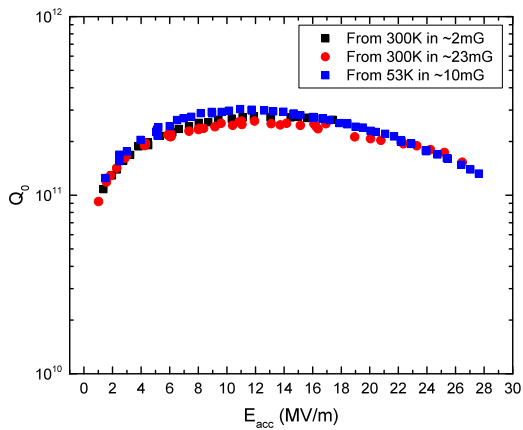}
\caption{\label{fig:RecordQ}$Q(E_\mathrm{acc}$) curves for three different cooldowns in different $B_\mathrm{avg}$.}
\end{figure}

First set of record results is shown in Fig.~\ref{fig:RecordQ} with the $Q$ of the cavity reaching above $2\times10^{11}$ up to the accelerating field of 22~MV/m for three different cooldowns in three different $B_\mathrm{avg}$ = 2 mG; 10 mG; 23 mG. No degradation in $Q$ with increasing ambient magnetic fields suggests that they are fully expelled. Furthermore, this full expulsion can be achieved by cooling from different starting temperatures.

\begin{figure}[htb]
 \includegraphics[width=\linewidth]{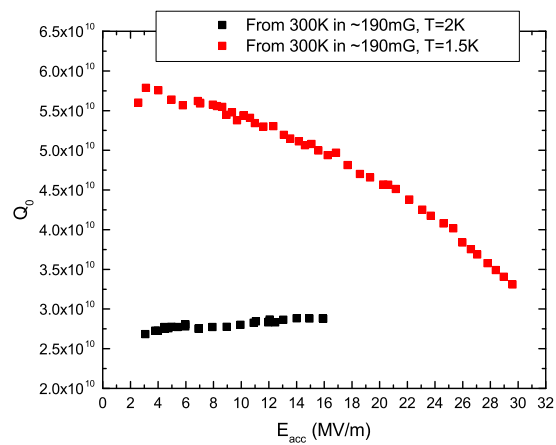}
\caption{\label{fig:190mG}$Q(E_\mathrm{acc}$) curves at $T=2$~K (black) and $T=1.5$~K measured after cooldown from 300~K in $B_\mathrm{avg}\approx$190~mG.}
\end{figure}

The second record result is shown in Fig.~\ref{fig:190mG}, and was obtained after cooling in $B_\mathrm{avg}\approx190$~mG, which is within a factor of two of the Earth magnetic field values. Yet the measured $Q=2.9\times10^{10}$ at 2~K and 16 MV/m is still high enough to satisfy the requirements of the LCLS-II project ($Q=2.7\times10^{10}$ at 16~MV/m), which has the highest $Q$ specification out of all SRF-based accelerators ever proposed or built.  

\begin{figure}[htb]
 \includegraphics[width=\linewidth]{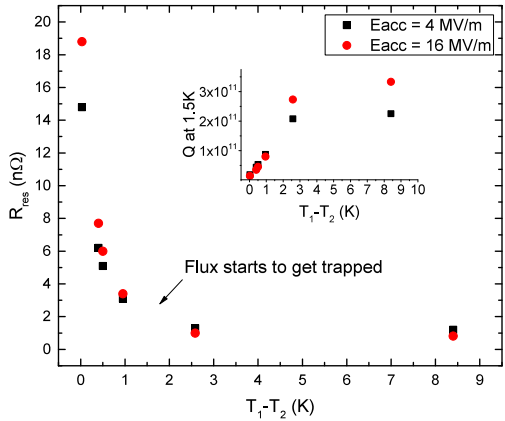}
\caption{\label{fig:Rres_DeltaT}Residual resistance at 1.5~K as a function of the temperature gradient present at the normal/superconducting front as measured when equator reaches $T_\mathrm{c}$. Inset shows the corresponding Q values.}
\end{figure}

\begin{figure}[htb]
 \includegraphics[width=\linewidth]{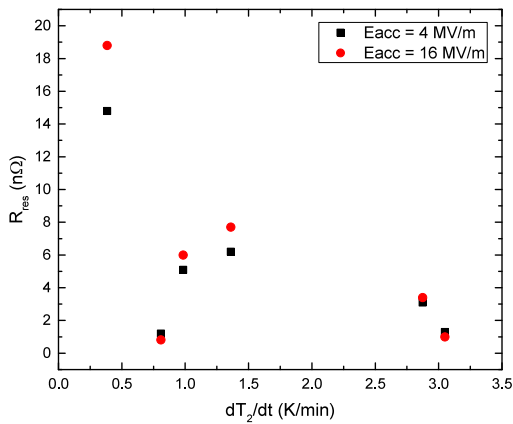}
\caption{\label{fig:Rres_dT_dt}Residual resistance at 1.5~K as a function of the cooling rate measured when equator is reaching $T_\mathrm{c}$.}
\end{figure}

In order to pinpoint the required thermal conditions for efficient flux expulsion, we have fixed the ambient field to 10~mG and varied temperature distribution along the cavity during cooling cycles by changing the starting temperature and helium flow rate. Obtained values of $R_\mathrm{res}$ are shown in Fig.~\ref{fig:Rres_DeltaT} (and $Q$ values in the inset) as a function of temperature difference between the top iris ($T_1$) and equator ($T_2$), and in Fig.~\ref{fig:Rres_dT_dt} as a function of the cooling rate $dT_2/dt | T_2=T_\mathrm{c}$. As it can be clearly seen, temperature difference at the phase front is the main factor for flux expulsion, while cooling rate itself has no clear effect. This finding is consistent with one of our proposed interpretations in~\cite{Romanenko_JAP_2014}.

\begin{figure}[htb]
 \includegraphics[width=\linewidth]{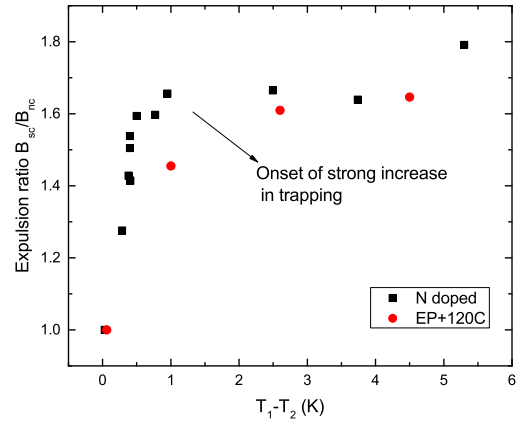}
\caption{\label{fig:Expulsion}Ratio of the magnetic field at the equator in superconducting state ($B_\mathrm{sc}$) to that in the normal state ($B_\mathrm{nc}$).}
\end{figure}

In our previous work~\cite{Romanenko_JAP_2014} we have shown via magnetostatic simulations that if the magnetic flux is fully expelled then the vertical component of the magnetic field at the equator should be increased by close to a factor of 1.8. Therefore, a ratio of the flux magnetometer readings before and right after the transition provides a measure of the amount of the flux expelled. In Fig.~\ref{fig:Expulsion} a summary plot for all the cooling procedures in various magnetic fields is shown. Fast increase in the trapped fraction (decrease in the expulsion ratio) is clearly observed as soon as the temperature difference across the top half-cell drops below $\sim$1-2~K, which corresponds to the gradient of 0.1-0.2~K/cm along the cavity surface. This increase in trapping causes the increase in $R_\mathrm{res}$ shown in Fig.~\ref{fig:Rres_dT_dt}.

An identical TESLA shape cavity but prepared by electropolishing/120$^\circ$C baking was measured in some cooldowns as well and exhibited a very similar qualitative behavior (red circles in Fig.~\ref{fig:Expulsion}). This suggests that the expulsion efficiency is not  determined by surface pinning properties, as 120$^\circ$C baked cavities have a drastically lower electron mean free path $\ell$ at the surface (and therefore different pinning strength)~\cite{Romanenko_LEM_APL_2014}.

In this paper we have shown that optimized Meissner expulsion procedure allows to completely eliminate the magnetic flux contribution and results in ultralow residual resistances even in high magnetic fields of up to 190~mG. If coupled with the ultralow BCS and non-flux residual resistance achieved via nitrogen doping, record quality factors of $> 2\times10^{11}$ emerge up to high fields. As one of the immediate practical implications, a variety of large-scale SRF-based projects , i.e. LCLS-II at SLAC, PIP-II at FNAL, can have a significantly lower operational power even with poor magnetic shielding. 

The implications also extend to any other superconducting devices involving trapped flux, where changing the temperature gradient during the transition through $T_\mathrm{c}$ can allow to tune the trapped flux amount for the fixed applied magnetic field.

Authors would like to acknowledge technical assistance of A. Rowe, M. Merio, B. Golden, Y. Pischalnikov, B. Squires, G. Kirschbaum, D. Marks, and R. Ward for cavity preparation and testing. Fermilab is operated by Fermi Research Alliance, LLC under Contract No. DE-AC02-07CH11359 with the United States Department of Energy.

%

\end{document}